\documentclass[12pt]{article}
\usepackage{amsbsy, amstext, amssymb, amsthm, amsmath,bm,bbm,wasysym}
\usepackage{graphicx,booktabs,hyperref,rotating,enumitem}
\usepackage{array,stmaryrd}
\usepackage{tabularx}
\usepackage{pdflscape}
\usepackage{adjustbox}
\usepackage{subfig}
\usepackage{longtable}
\usepackage{listings}
\usepackage{multirow}
\usepackage{xr}
\usepackage{blkarray}
\usepackage{setspace}
\usepackage{algorithm}
\usepackage{algpseudocode}

\usepackage[natbibapa]{apacite}
\usepackage{etoolbox}
\patchcmd{\APACjournalVolNumPages}{\unskip({#3})}{}{}{} 
\patchcmd{\APACjournalVolNumPages}{\Bem{#2}}{#2}{}{}

\AtBeginEnvironment{APACrefURL}{\renewcommand{\url}[1]{}}
\AtBeginDocument{}
\AtBeginEnvironment{longtable}{\singlespacing}

\renewcommand{\baselinestretch}{1.2}
\usepackage{tikz}
\usetikzlibrary{shapes.geometric, arrows}
\usetikzlibrary{arrows.meta}
\usetikzlibrary{shapes,backgrounds,calc,positioning}

\renewcommand\footnoterule{\kern-3pt \hrule \textwidth 2in \kern 2.6pt}
\def\authorfootnote#1{{\let\thefootnote\relax\footnotetext{#1}}}

\newcommand{\E}{\mathop{\mathbb{E}}} 

\newcommand{\real}[1]{\mathbb{R} \mathit{^{#1}}}

\newcommand{\Xbf}{{\bm X}}

\newcommand{\betabf}{\boldsymbol{\beta}}

\usepackage{comment}

\begin{document}
\def\spacingset#1{\renewcommand{\baselinestretch}%
{#1}\small\normalsize} \spacingset{1}

\thispagestyle{empty}
\baselineskip=28pt
\title{\bf Prediction Inference Using Generalized Functional Mixed Effects Models}
\author{Xinkai Zhou$^{*,1}$, Erjia Cui$^{*,2}$, Joseph Sartini$^{1}$, Ciprian Crainiceanu$^{1}$\\
  \small{$^{1}$Department of Biostatistics, Johns Hopkins University} \\
  \small{$^{2}$Division of Biostatistics and Health Data Science, University of Minnesota} \\
  \small{The * indicates co-first authors}}
  \date{}
\maketitle

\bigskip
\begin{abstract}
We introduce inferential methods for prediction based on functional random effects in generalized functional mixed effects models. This is similar to the inference for random effects in generalized linear mixed effects models (GLMMs), but for functional instead of scalar outcomes. The method combines: (1) local GLMMs to extract initial estimators of the functional random components on the linear predictor scale; (2) structural functional principal components analysis (SFPCA) for dimension reduction; and (3) global Bayesian multilevel model conditional on the eigenfunctions for inference on the functional random effects. Extensive simulations demonstrate excellent coverage properties of credible intervals for the functional random effects in a variety of scenarios and for different data sizes. To our knowledge, this is the first time such simulations are conducted and reported, likely because prediction inference was not viewed as a priority and existing methods are too slow to calculate coverage. Methods are implemented in a reproducible \texttt{R} package and demonstrated using the NHANES 2011-2014 accelerometry data.
\end{abstract}

\noindent%
\vfill

\newpage
\spacingset{1.25} 
\section{Introduction} \label{sec:intro}

We introduce a class of inferential methods to quantify the prediction uncertainty in generalized (Gaussian and non-Gaussian) functional mixed effects models. Specifically, our goal is to construct nominal credible intervals for functional random effects on the linear predictor scale. This framework extends classical  random effects inference methods in generalized linear mixed effects models (GLMMs) from scalar to functional outcomes. Such methods are becoming increasingly essential in studies that collect individual-level data and focus on predicting individual outcomes rather than characterizing population-level quantities.

Consider, for example, the scenario where physical activity is measured using accelerometers every minute over an extended period (e.g., weeks or months). Such data structure raises many questions, including: (1) Are measurements from a particular individual  unusual compared to those of others? (2) For a particular individual, when and where do new measurements deviate from their historical records? (3) How to quantify the uncertainty of these assessments? All these questions can be framed as inferential problems involving functional random effects.

\begin{figure}[!tbh]
  \centering
  \includegraphics[width=\textwidth]{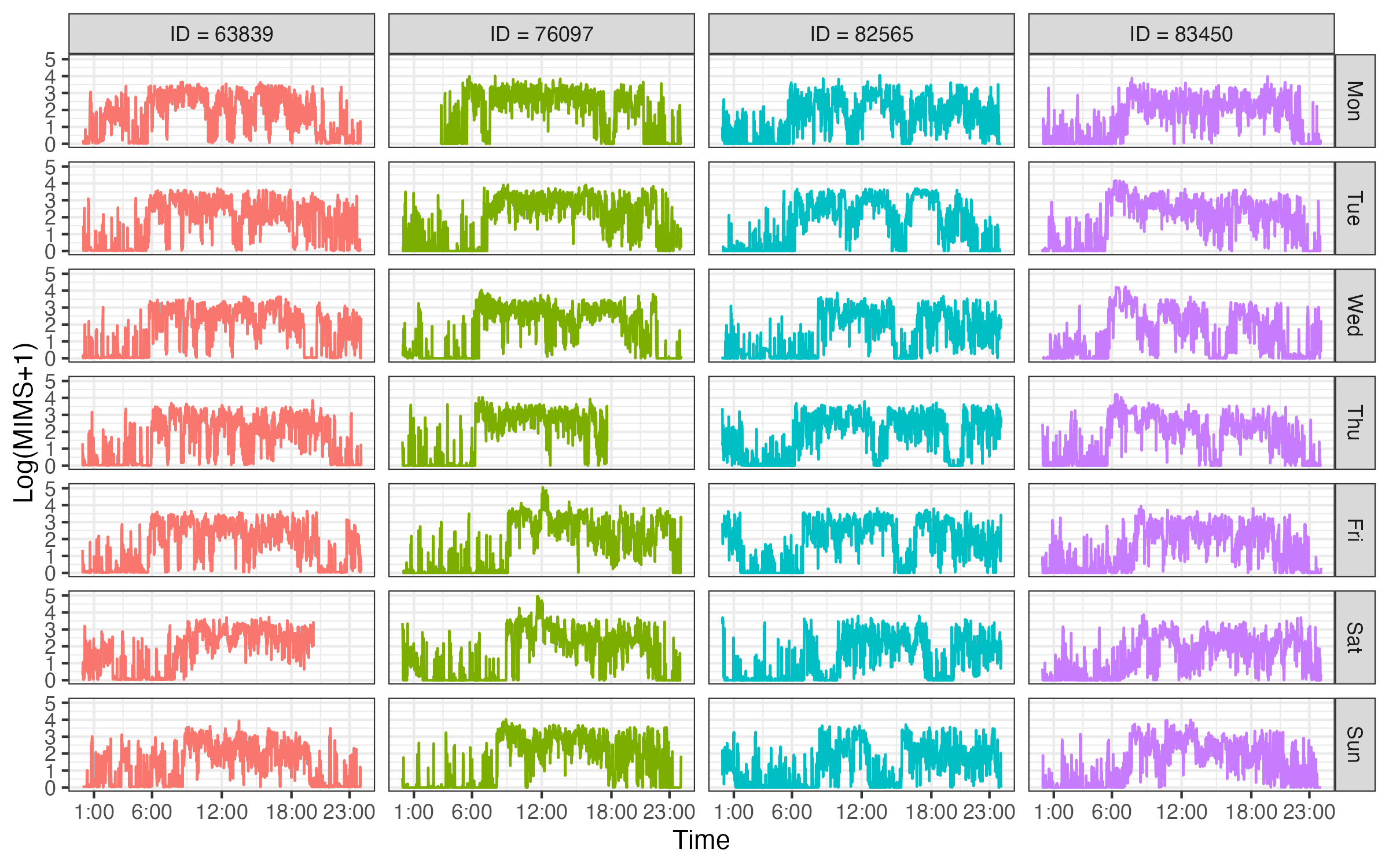}
  \caption{Minute-level objective physical activity data in log(MIMS + 1) units for four NHANES study participants. Each column represents one participant. Each row represents one day of the week from Monday to Sunday. Each panel displays $1{,}440$ observations from midnight to midnight.}
  \label{fig:figure1}
\end{figure}

Our work is motivated by the accelerometry data collected as part of the 2011-2014 National Health and Nutrition Examination Survey (NHANES). Specifically, $14{,}693$ participants were asked to wear a wrist-worn accelerometer for seven consecutive days to continuously monitor their physical activity. The accelerometry data was extracted, processed and released as minute-level Monitor-Independent Movement Summary (MIMS), a unit that quantifies physical activity intensity \citep{john2019mims}. Figure \ref{fig:figure1} displays minute-level, log-transformed MIMS values for four randomly selected NHANES participants. Each column represents one study participant, and each row represents one day of the week from Monday to Sunday. Upon examining Figure~\ref{fig:figure1}, several features become evident: (1) physical activity patterns vary substantially across subjects; (2) within a subject, physical activity patterns can be similar or different across days; (3) missing data are not uncommon and they tend to occur in contiguous windows (for example, from 6pm to 12am on Thursday for subject ID 76097). Given the substantial heterogeneity and potential missingness in physical activity trajectories, determining how to effectively decompose and quantify such variability becomes an important but challenging problem. 

The NHANES accelerometry data is an example of multilevel high-dimensional data, because minute-level MIMS data (high-dimensional) was collected over multiple days of the week for each subject (multilevel). Data sets that contain complex dependencies between functional observations, labeled ``second-generation" functional data, is becoming increasingly common and requires tailored modeling techniques \citep{koner_second-generation_2023}. Examples of these interdependence structures include spatial  \citep{staicu2010fast,li2021multilevel,qian2024multivariate, burbano-moreno_spatial_2024}, temporal \citep{greven2011longitudinal, park_longitudinal_2015, shamshoian_bayesian_2020, alam_modeling_2024, sergazinov2023case, zipunnikov2014longitudinal, zhu2019fmem}, nested/crossed \citep{serban_multilevel_2013, shou2015structured, goldsmith2015generalized, brockhaus2015functional}, latent clustering \citep{marco_functional_2024}, and multivariate  \citep{gunning2024mmlf,lin2024multilevel, cao2024efficient, volkmann_multivariate_2023} designs. These data structures can be modeled as functional responses with scalar predictors, where the correlation induced by the sampling mechanism is modeled via structured functional residuals. The functional mixed effects models \citep{guo2002functional, morris2006wavelet, goldsmith_penalized_2011, scheipl2015functional, scheipl2016generalized, greven_general_2017, cui2022fui, sun2023ultra} is a widely used framework to represent these dependency structures. Despite the extensive methodological and applied research in this area, studies focusing on functional random effects inference and, in particular, on quantifying the performance of these methods remain relatively sparse. For an in-depth introduction to these models see, for example, Chapter 8 in \cite{fda_with_R}.

The present work focuses on constructing credible intervals for the functional random effects and quantifying their statistical properties. In a previous study, \cite{cao2018robust} proposed a method for subject-specific inference in the single-level setting, but reported under-coverage of their confidence intervals. Surprisingly, the paper by \cite{cao2018robust} is the only paper that we could find that addressed the performance of credible intervals for functional random effects. This could be due to the computational challenges associated with high dimensionality and fitting functional mixed effects models. Furthermore, the existing literature does not address missing data and does not scale up to large data sets, such as our motivating NHANES application.

To address this critical gap, we introduce the Functional Random-effect Inference Method (FRIM).  FRIM produces credible intervals for subject- and subject-visit level functional random effects while allowing for (1) missing data, (2) regularly or irregularly spaced observations, and (3) generalized (Gaussian or non-Gaussian) functional data. Because FRIM is built on the idea of scalability, methods can be used for increasingly large data sets, and simulations can be conducted at medium to large sample sizes to study the properties of credible intervals. The method is accompanied by open-source \texttt{R} software to ensure reproducibility. 

The rest of the paper is organized as follows. 
In Section~\ref{sec:methods}, we introduce the FRIM framework. Simulation results are presented in Section~\ref{sec:sim}. The NHANES application studies are described in Section~\ref{sec:app}. We conclude with discussions in Section~\ref{sec:discussion}.

\section{Functional Random Effect Inference}\label{sec:methods}

The Functional Random-effect Inference Method (FRIM) framework consists of the following steps: (1) use local GLMMs to extract initial estimates of the functional random components on the linear predictor scale; (2) apply structured functional principal components (SFPCA) \citep{shou2015structured} for dimension reduction; and (3) fit a global Bayesian multilevel model conditional on the eigenfunctions for functional random effects inference. Each step is explained in detail below.

\subsection{Binning and Local Estimates}\label{subsubsec:binning}

The observed data are of the form $\{\Xbf_{ij}, (s_{ijl}, Y_{ijl})_{l=1}^{L_{ij}}\}$,  where $\mathbf{X}_{ij} = [X_{ij1}, X_{ij2}, \ldots, X_{ijp}]^T \in \real{p}$ is a vector of fixed effect covariates for study participant $i$ ($i=1,\ldots, I$) at visit $j$ ($j=1,\ldots,J_i$), and should include at least an intercept and dummy variables indicating visit. $Y_{ijl}$ is the functional observation at location $s_{ijl}\in S$, where  $l = 1, \ldots, L_{ij}$. This data structure does not require $Y_{ijl}$ to be observed on the same grid across visits and/or subjects, which is an important distinction from the structure assumed in \citet{cui2022fui} and \citet{sun2023ultra}. 

Let $\{c_1, \ldots, c_M\} \in S$ be the bin centers. Given a bin width $d$, the bin $\mathcal{C}_m$ 
centered at $c_m$ consists of all data points within distance $\frac d 2$ from $c_m$, i.e., $\mathcal{C}_m = \{(s_{ijl}, Y_{ijl}): |s_{ijl} - c_m| \leq \frac d 2 \}$. To estimate local fixed and random effects, the following GLMM is fit in each data bin $\mathcal{C}_m$:
\begin{equation}
\begin{aligned}\label{eq:local-glmm}
    \eta_{ij}(c_m) & = g\big\{E\big(Y_{ijl}\mid a_i(c_m), b_{ij}(c_m), Y_{ijl} \in \mathcal{C}_m \big) \big\}  = \mathbf{X}_{ij}\boldsymbol{\beta}(c_m) +  
    a_i(c_m) + b_{ij}(c_m)\;,
\end{aligned}
\end{equation}
where $a_i(c_m)$ and $b_{ij}(c_m)$ are subject- and subject-visit-specific random intercepts, respectively. Notice that the method can be applied to any type of generalized (Gaussian or non-Gaussian) functional outcomes. With a slight abuse of notation, we use the bin center $c_m$ in the parentheses to indicate quantities that correspond to the local GLMM model for the bin $\mathcal{C}_m$; it is worth emphasizing that $\boldsymbol{\beta}(c_m)$, $a_i(c_m)$, and $b_{ij}(c_m)$ are not functions, because they are estimated over a discrete set of bins. Denote the estimated fixed effect as $\widehat{\boldsymbol{\beta}}(c_m) = \{\widehat\beta_1(c_m), \ldots, \widehat\beta_p(c_m)\}^T$, the estimated random effect as $\widehat{r}_{ij}(c_m) = \widehat{a}_i(c_m) + \widehat{b}_{ij}(c_m)$, and the estimated linear predictor as $\widehat{\eta}_{ij}(c_m)$. 

From a statistical perspective, binning offers several key advantages. First, it enables the prediction of subject-visit-specific random effect $b_{ij}(c_m)$, which is only estimable when multiple observations are available for each subject-visit pair and therefore cannot be estimated using pointwise models as in fast univariate inference (FUI) \citep{cui2022fui}. Second, binning borrows information from nearby points, improving estimation stability. This is especially important for distributions where extreme data imbalance (e.g., too many zeros in binary data) at certain locations may cause convergence issues for pointwise fitting. Third, binning naturally accommodates irregular observations, allowing locations of observations $\{s_{ijl}\}$ to differ across subjects and/or visits. While the concept of binning has been explored before \citep{leroux2023gfpca, zhou2023gmfpca, lu2024gcfpca}, inference for functional random effects has not.

\subsection{Random Component Estimation and Decomposition}\label{subsubsec:random}

After obtaining the local fixed effects estimates $\widehat{\boldsymbol{\beta}}(c_1), \ldots, \widehat{\boldsymbol{\beta}}(c_M)$ from each bin, the functional fixed effects estimates $\widetilde{\boldsymbol{\beta}}(s) = [\widetilde{\beta}_1(s), \widetilde{\beta}_2(s), \ldots, \widetilde{\beta}_p(s)]^T$ are derived by smoothing the $M$ local estimates for each of the $p$ predictors separately along the functional domain. The fixed component of the model is then estimated by $\Xbf_{ij}\widetilde\betabf(s)$. This two-step procedure of local fitting and smoothing was first proposed in FUI and was shown to achieve accurate estimation of functional fixed effects through extensive simulations.

Since the functional fixed effects estimates are smoothed, the random effects estimates need to be adjusted accordingly. We calculate the adjusted random component estimate $\widehat{r}_{ij}^a(\cdot)$ by subtracting the smoothed fixed component estimate $\Xbf_{ij}\widetilde\betabf(\cdot)$ from the locally estimated linear predictor $\widehat\eta_{ij}(\cdot)$ at the center of each bin. Specifically, 
\begin{equation*}
\widehat{r}_{ij}^a(c_m) = \widehat\eta_{ij}(c_m) - \Xbf_{ij}\widetilde\betabf(c_m) = \hat{r}_{ij}(c_m) + \Xbf_{ij}[\widehat\betabf(c_m) - \widetilde\betabf(c_m)]\;, 
\end{equation*}
for $m = 1, \ldots, M$. Calculating $\widehat{r}_{ij}^a(\cdot)$ is an important step because it preserves the smoothness of functional fixed effects estimates while minimizing the number of parameters to smooth. Indeed, by smoothing only the $p$ fixed effects coefficients instead of all linear predictors as in FUI, the number of parameters to smooth remains constant as sample size increases, making the approach more efficient for larger data sets. 

Given the multilevel functional structure of $\widehat{r}_{ij}^a(c_m)$, multilevel FPCA \citep{di2009multilevel}, which is a special type of structured FPCA \citep{shou2015structured}, serves as a powerful tool for variance decomposition. Specifically, $\widehat{r}_{ij}^a(s)$ can be decomposed as
\begin{equation}
    \label{eqn:frim-mfpca}
    \widehat{r}_{ij}^a(s) = \sum_{k_1 = 1}^{\infty} \xi_{ik_1} \phi_{k_1}(s) + \sum_{k_2 = 1}^{\infty} \zeta_{ijk_2} \psi_{k_2}(s) + \epsilon_{ij}(s)\;,
\end{equation}
where $\phi_{k_1}(\cdot)$ is the $k_1$-th subject-level eigenfunction that characterizes the $k_1$-th dominant direction of variation on the subject-level, $\xi_{ik_1} \sim \mathcal{N}(0, \lambda_{k_1}^{(1)})$ is the corresponding score for subject $i$, and $\lambda_{k_1}^{(1)}$ is the corresponding eigenvalue. Similarly, $\psi_{k_2}(\cdot)$ and $\zeta_{ijk_2}\sim \mathcal{N}(0, \lambda_{k_2}^{(2)})$ are subject-visit-level counterparts. The noise term $\epsilon_{ij}(s)$ accounts for errors induced by using local estimates and is the reason why local random effects estimates $\widehat{r}_{ij}^a(c_m) ~ (m = 1, \ldots, M)$ do not need to be smoothed. Model~\eqref{eqn:frim-mfpca} does not contain fixed effects, because the fixed effects were extracted in model~\eqref{eq:local-glmm}. In practice, we use fast MFPCA \citep{cui2023mfpca} implemented in the \texttt{mfpca.face()} function of the \texttt{R} package \texttt{refund}. This approach substantially improves the computational performance using the fast covariance estimation introduced in \cite{xiao2016fast}. Dimension reduction is achieved by keeping the top $K_1$ eigenfunctions in level one and top $K_2$ eigenfunctions in level two, where $K_1$ and $K_2$ are chosen to ensure that high proportions of variance are explained at each level. The smoothed and dimension-reduced approximation to the adjusted random component estimates from model \eqref{eqn:frim-mfpca}, denoted as $\widehat{r}_{ij}^d(s)$, becomes 
\begin{align}\label{eq:mfpca-thresholded}
    \widehat{r}_{ij}^d(s) = \sum_{k_1 = 1}^{K_1} \xi_{ik_1} \phi_{k_1}(s) + \sum_{k_2 = 1}^{K_2} \zeta_{ijk_2} \psi_{k_2}(s)\;. 
\end{align}

\subsection{Score Sampling Using a Bayesian Multilevel Model} \label{sec:step4}

Structured FPCA provides substantial dimensionality reduction while preserving the multilevel structure of the data, but is not designed to quantify the uncertainty of functional random effects. We solve this problem by using Bayesian inference by conditioning on the estimated eigenfunctions $\{\hat{\phi}_{k_1}(s)\}$, $k_1 = 1, \ldots, K_1$ and $\{\hat{\psi}_{k_2}(s)\}$, $k_2 = 1, \ldots, K_2$, of the hierarchy. The model is
\begin{equation}
\begin{aligned}\label{eq:bayeshier}
    & g\Big(\E\big(Y_{ij}(s) \mid \widetilde\betabf(s), \{\widehat{\phi}_{k_1}(s)\}_{k_1 = 1}^{K_1}, \{\widehat{\psi}_{k_2}(s)\}_{k_2 = 1}^{K_2} \big)\Big)\\
    & \;\;\;\;\;\;\;\; = \Xbf_{ij}\tilde\betabf(s) + \sum_{k_1 = 1}^{K_1} \xi_{ik_1} \widehat{\phi}_{k_1}(s) + \sum_{k_2 = 1}^{K_2} \zeta_{ijk_2} \widehat{\psi}_{k_2}(s)\;,
\end{aligned}
\end{equation}
where standard priors for Bayesian multilevel models were used, i.e., $\xi_{ik_1}\sim \mathcal{N}(0, \sigma^2_{\xi_{k_1}})$, $\zeta_{ijk_2} \;\sim \mathcal{N}(0, \sigma^2_{\zeta_{k_2}})$ are mutually independent random variables and the inverse variance components $1/\sigma^2_{\xi_{k_1}}$ and $1/\sigma^2_{\zeta_{k_2}}$ were assigned independent $\text{Cauchy}(0, 1)$ priors. 
Here we treated $\widetilde\betabf(s)$ as fixed for computational efficiency, but a fully Bayesian analysis of this model could be implemented as well. The estimated random component from the $b$-th MCMC iteration is $\widetilde{r}_{ij}^{(b)}(s_{ijl}) = \sum_{k_1 = 1}^{K_1} \xi_{ik_1}^{(b)} \widehat\phi_{k_1}(s_{ijl}) + \sum_{k_2 = 1}^{K_2} \zeta_{ijk_2}^{(b)} \widehat\psi_{k_2}(s_{ijl})$.

Finally, the $95$\% credible intervals for subject-visit-specific random effect can be constructed using $\{\widetilde{r}_{ij}^{(b)}(s_{ijl}), b = 1, \ldots, B\}$, where $B$ is the total number of MCMC iterations.  

Compared to the fully Bayesian approach proposed by \cite{goldsmith2015generalized}, FRIM improves computational efficiency by conditioning on the eigenfunctions, thereby substantially reducing the dimensionality of the problem. Additionally, the method is easy to use and can be readily implemented using existing statistical software such as \texttt{Stan} \citep{carpenter2017stan}. 

\subsection{Algorithm}

We summarize the steps of FRIM in Algorithm \ref{algo:frim}.  The implementation can be found in the \texttt{R} package \texttt{FRIM} (\url{https://github.com/xinkai-zhou/frim}).

\begin{algorithm}[H]
  1. Select $M$ bin centers $c_1, \ldots, c_M$ and assign data points to corresponding bins. For each bin, fit a local GLMM using data from that bin. Obtain local linear predictor estimates $\widehat{\eta}_{ij}(c_m)$ and local fixed effect estimates $\widehat{\boldsymbol{\beta}}(c_m)$. \\
  2. Smooth $\widehat{\boldsymbol{\beta}}(c_m)$ to obtain functional fixed effects estimate $\widetilde{\boldsymbol{\beta}}(s)$. Calculate the adjusted random component estimates $\widehat{r}_{ij}^a(s)$. Perform fast MFPCA on $\widehat{r}_{ij}^a(s)$ to estimate the eigenfunctions $\widehat{\phi}_{k_1}(s)$ and $\widehat{\psi}_{k_2}(s)$. \\
  3. Fit a Bayesian multilevel model on the full data to obtain the posterior distributions of scores and functional random effects conditional on the eigenfunctions.
  \caption{Functional Random-effect Inference Method (FRIM)}
  \label{algo:frim}
\end{algorithm}

\subsection{Missing Data}
\label{sec:missing-data}
The ability to handle missing data is crucial for real world applications. For instance, the NHANES accelerometer data were collected in the free-living environment and missingness occurred frequently due to battery depletion, device malfunction, or non-wear. FRIM handles missing data implicitly. Indeed, consider bin $\mathcal{C}_m = \{(s_{ijl}, Y_{ijl}): |s_{ijl} - c_m| \leq \frac d 2 \}$ and consider the case of partial missing data. More precisely,  for subject $i$ and visit $j$, assume that some, but not all, $Y_{ijl}$ are missing in $\mathcal{C}_m$. The local random effect $\widehat{r}_{ij}(c_m)$ is estimable using the GLMM inferential machinery. When  all $Y_{ijl}$ are missing for subject $i$ and visit $j$ in bin $\mathcal{C}_m$, the local random effect $\widehat{r}_{ij}(c_m)$ is not estimable and leads to missing values in $\widehat{r}_{ij}^a(c_m)$. MFPCA, however, can handle such missing data patterns when the resulting missing data is interspersed with observed data \citep{xiao2016fast, cui2023mfpca}. 

On the other hand, when data are missing in large contiguous blocks for a sizable number of study participants, simulation experiments suggest that MFPCA may not perform as well in terms of recovering eigenfunctions that correspond to small variances. Our strategy for handling such cases is to exclude visits that contain missing data when estimating eigenfunctions to ensure that eigenfunctions are consistently estimated, which works well when data are missing at random. Once eigenfunctions are estimated, the Bayesian multilevel model can be fit on the original data set, including days with substantial missingness. We will show in Section \ref{sec:sim} that this strategy works well for both Gaussian and binary data with various missing rates. In summary, FRIM is robust to missing data, and its divide-and-conquer approach provides multiple points for testing assumptions and refining models. 

\subsection{Information Leakage between Random and Fixed Effects}
\label{sec:leakage}
In simulations, we noticed that when the number of subjects $I$ is small, the fixed effect estimate for the visit-$j$ effect may contain patterns that resemble a linear combination of level-two eigenfunctions. To better understand this phenomenon, consider, without loss of generality, that the data is Gaussian and follows the model
\begin{align*}
    Y_{ij}(s) = \beta_0(s) + \sum_{k_1 = 1}^{K_1} \xi_{ik_1} \phi_{k_1}(s) + \sum_{k_2 = 1}^{K_2} \zeta_{ijk_2} \psi_{k_2}(s) + \epsilon_{ij}(s)\;,
\end{align*}
with all the standard assumptions. The population mean over all subjects and visits is 
\begin{eqnarray*}
    \overline Y_{..}(s) = \beta_0(s) + \sum_{k_1=1}^{K_1}\frac{\sum_{i=1}^I \xi_{ik_1}}{I} \phi_{k_1}(s) + \sum_{k_2 = 1}^{K_2} \frac{\sum_{i=1}^I\sum_{j=1}^J \zeta_{ijk_2}}{I\times J} \psi_{k_2}(s) + \frac{\sum_{i=1}^I\sum_{j=1}^J \epsilon_{ij}(s)}{I \times J}\;.
\end{eqnarray*}
The visit-specific mean for the $j$-th visit is 
\begin{eqnarray*}
    \overline Y_{.j}(s) = \beta_0(s) + \sum_{k_1=1}^{K_1}\frac{\sum_{i=1}^I \xi_{ik_1}}{I} \phi_{k_1}(s) + \sum_{k_2 = 1}^{K_2} \frac{\sum_{i=1}^I \zeta_{ijk_2}}{I} \psi_{k_2}(s) + \frac{\sum_{i=1}^I \epsilon_{ij}(s)}{I}\;.
\end{eqnarray*}
Therefore, the deviation of the $j$th visit from the population mean is 
\begin{eqnarray*}
    \overline Y_{.j}(s) - \overline Y_{..}(s) = \sum_{k_2 = 1}^{K_2} [\frac 1I \sum_{i=1}^I(\zeta_{ijk_2} - \frac{\sum_{j=1}^J \zeta_{ijk_2}}{J})] \psi_{k_2}(s) + \frac 1I \sum_{i=1}^I (\epsilon_{ij}(s) - \frac{\sum_{j=1}^J \epsilon_{ij}(s)}{J})\;.
\end{eqnarray*}
When $I$ is small, the term $\frac 1I \sum_{i=1}^I(\zeta_{ijk_2} - \frac{\sum_{j=1}^J \zeta_{ijk_2}}{J})$ can be noticeably different from zero. As a result, the estimate for the visit-$j$ deviation could display patterns that resemble a linear combination of the level-2 eigenfunctions, $\{\psi_{k_2}(s)\}_{k_2=1}^{K_2}$. We refer to this phenomenon as the \emph{Information Leakage} between random and fixed effects. In simulations, this phenomenon is observable for $I = 100$, which can negatively impact the inference for functional random effects. However,  $\frac 1I \sum_{i=1}^I(\zeta_{ijk_2} - \frac{\sum_{j=1}^J \zeta_{ijk_2}}{J})$ converges to zero as $I$ increases, resulting in an improved quality of inference for random effects as sample size increases.

\section{Simulations}\label{sec:sim}

\subsection{Simulation Setup}
We simulated Gaussian and binary multilevel functional data to evaluate the coverage rate of credible intervals for the random components produced by FRIM. The linear predictors  $\eta_{ij}(s_l)$ were generated from the following model:
\begin{eqnarray*}
\eta_{ij}(s_l) = X_i\beta(s_l) + \sum_{k_1 = 1}^{K_1} \xi_{ik_1} \phi_{k_1}(s_l) + \sum_{k_2 = 1}^{K_2} \zeta_{ijk_2} \psi_{k_2}(s_l)\;,
\end{eqnarray*}
where $\{s_l = l/L: l = 1, \ldots, L\}$ are sampling locations on the functional domain and $L$ is the number of sampling points. The fixed component only included an intercept term because the estimation and inference of fixed effects through a two-step procedure is relatively well understood through extensive simulation experiments in \citet{cui2022fui}. Furthermore, the inference of fixed effects is not the focus of this paper. The corresponding functional fixed effect was set to $\beta(s) = \sqrt{2} \sin(2\pi s)$. Scores were generated as $\xi_{ik_1} \sim \mathcal{N}(0, \lambda_{k_1})$ and $\zeta_{ijk_2} \sim \mathcal{N}(0, \lambda_{k_2})$, where the true eigenvalues were $\lambda_{k_1} = 0.5^{k_1-1}$ for $k_1 = 1, 2, ..., K_1$ and $\lambda_{k_2} = 0.5^{k_2-1}$ for $k_2 = 1, 2, ..., K_2$. We used $K_1 = K_2 = 4$ throughout simulation experiments.  For eigenfunctions, we considered the following two cases:

\noindent\textit{Case 1.} Mutually orthogonal bases.
\begin{enumerate}[label=Level \arabic*:, align=left, leftmargin=1in]
    \item [Level 1:] $\phi_l(s) = \{\sqrt{2} \sin(4\pi s), \sqrt{2} \cos(4\pi s), \sqrt{2} \sin(6\pi s), \sqrt{2} \cos(6\pi s)\}\;.$
    \item [Level 2:] $\psi_m(s) = \{\sqrt{2} \sin(8\pi s), \sqrt{2} \cos(8\pi s), \sqrt{2} \sin(10\pi s), \sqrt{2} \cos(10\pi s)\}\;.$
\end{enumerate}
\noindent\textit{Case 2.} Mutually orthogonal within each level, but not orthogonal between levels.
\begin{enumerate}[label=Level \arabic*:, align=left, leftmargin=1in]
    \item same as Case 1.
    \item $\psi_1(s) = 1, \psi_2(s) = \sqrt 3 (2s-1), \psi_3(s) = \sqrt 5 (6s^2-6s+1), \psi_4(s) = \sqrt 7 (20s^3 - 30s^2 + 12s - 1)$.
\end{enumerate}
 Gaussian functional data were generated as $Y_{ij}(s_l) \sim N(\eta_{ij}(s_l), \sigma^2_\epsilon)$ where $\sigma^2_\epsilon = 1$, and binary data were generated as $Y_{ij}(s_l) \sim \text{Bernoulli}\{\mu_{ij}(s_l)\}$, where ${\rm logit}\{\mu_{ij}(s_l)\}= \eta_{ij}(s_l)$. In addition to varying the eigenfunctions, we also examined the effect of (1) number of subjects $I=100, 500, 1000$; (2) number of visits per subject $J = 2, 5, 10$; 
and (3) bin widths that correspond to $W=2\%, 5\%, 10\%$ of the number of sampling points. We varied one simulation parameter at a time while keeping the others fixed. 

To evaluate how FRIM performs in the presence of missingness, we simulated missing data through a two-step procedure: first, sample an indicator variable $M_{ij} \sim \text{Bernoulli}(0.2)$ for each subject-visit pair indicating whether the functional observation $Y_{ij}(\cdot)$ contains missing data; second, for subject-visit pairs with $M_{ij} = 1$, randomly convert $M = 10\%, 25\%, 50\%$ of the observations to missing. The missingness was induced in contiguous blocks to mimic the missing pattern of the real accelerometry data. Specifically, we randomly sampled a left endpoint $s_a$ from $\{s_l = \frac{l}{L}: l = 1, ..., L\}$, calculated the right endpoint $s_b = s_a + M*L$, and set $Y_{ij}(s)$ to missing for $s_a \leq s < s_b$. 

In all experiments, we set the number of post warm-up iterations to $B = 2,000$ to ensure convergence of MCMC sampling. Each simulation scenario was repeated $100$ times where the simulated scores were fixed across repetitions.  

We report the mean pointwise coverage probability (MPCP) of the random components $r_{ij}(s_l)$. Recall from Section \ref{sec:step4} that we have access to the full posterior samples $\{\widetilde{r}_{ij}^{(b)}(s_l), b = 1, ..., B\}$, from which the posterior mean and $95$\% credible interval can be constructed. The coverage probability for $r_{ij}(s_l)$ can be estimated as the proportion of simulation replicates where the $95$\% credible interval covers the truth. Finally, MPCP is obtained by averaging coverage probabilities across subjects,  visits, and domain points. For experiments involving missing data, we report MPCP during periods of missingness. 

\subsection{Simulation Results}

\begin{figure}[!tbh]
\centering
\includegraphics[width=0.32\textwidth]{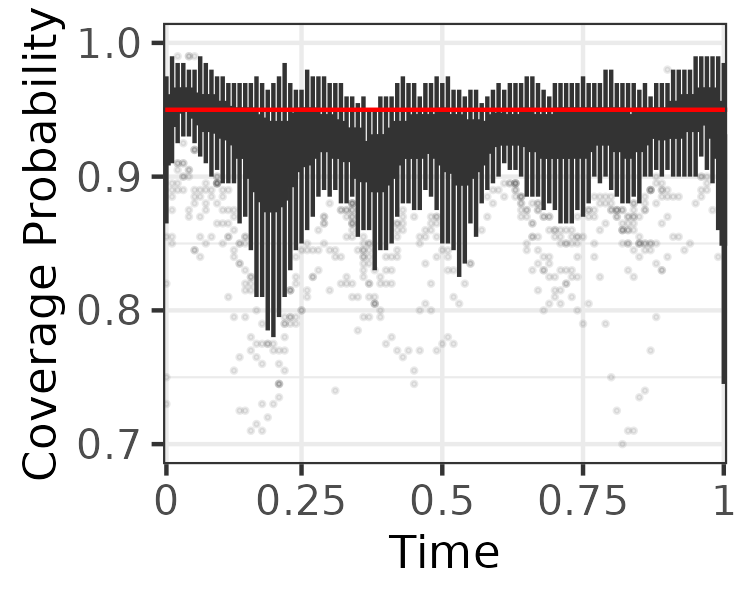}
\includegraphics[width=0.32\textwidth]{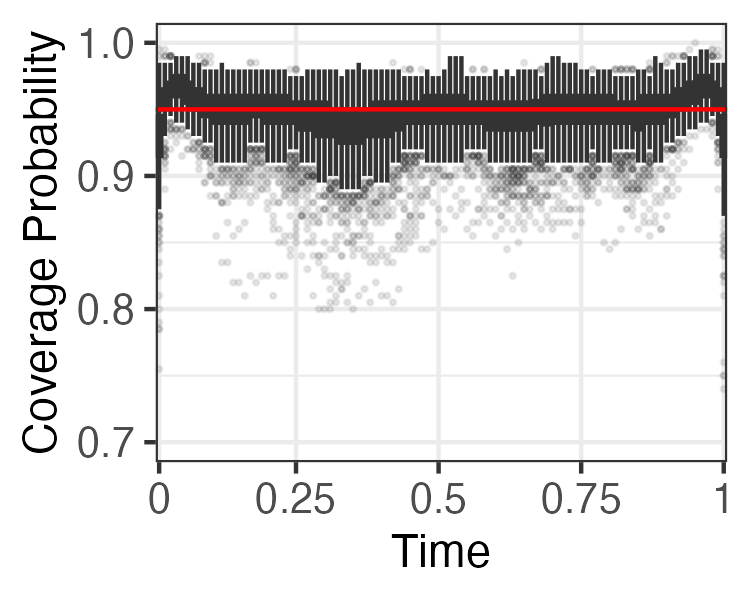}
\includegraphics[width=0.32\textwidth]{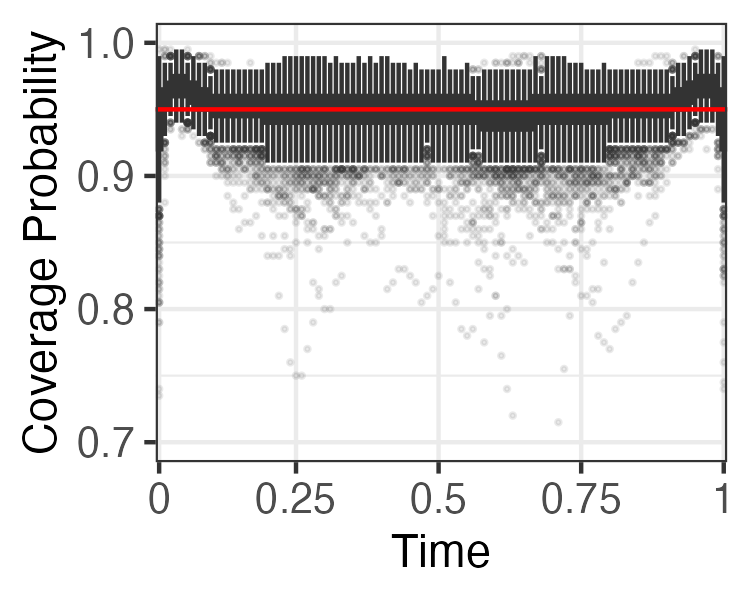}
  \caption{Boxplots of coverage probabilities averaged among visits from the same subject at each domain point. Left to right: sample size $I = 100, 500, 1000$. The experiment was based on binary data simulated using case 2 eigenfunctions, number of visits $J = 10$, number of sampling points $K = 100$, and percentage of data used for constructing local bins $w = 5\%$. The red horizontal line indicates the nominal level of 0.95. }
  \label{fig:covprob-ranef-by-I}
\end{figure}

First we show simulation results for complete data. We focus on the median (across visits within each subject) coverage probability of the $95\%$ credible intervals of the subject-visit-specific random effects from $100$ repeated experiments at each sampling point. Figure \ref{fig:covprob-ranef-by-I} displays the subject-specific coverage probabilities for binary data at each sampling point across different sample sizes $I$, where we used case 2 eigenfunctions and set the number of visits $J = 10$ and the number of sampling points $L = 100$. For $I=100$, the mean coverage hovers around $90$\% rather than the nominal level of $95$\%, which could be due to the information leakage phenomenon discussed in Section \ref{sec:leakage}. As the number of subjects increases, coverage quickly improves and centers around the nominal level of $95$\%. We also notice a slight degradation of coverage near either side of the domain boundary, which is likely a result of having fewer data points in bins near the boundary. 


Table \ref{table:mpcp-by-I} displays the MPCP statistics for Gaussian and binary data using both types of basis functions for different number of subjects $I$. In this experiment, we fixed the number of visits $J = 10$, number of sampling points $L = 100$, and the percentage of data used for constructing local bins $W$ to be $5\%$ of the data. Both Gaussian and binary data suffer from under-coverage when $I=100$. However, as sample size increases, the effect of random effects leakage gradually fades and the coverage quickly improves. At $I=1000$, coverage probabilities are close to the nominal level of $0.95$ for both types of data and basis functions. Results for different numbers of visits and bin widths and for computational efficiency can be found in Supplementary Materials Table \ref{supp-table:mpcp-by-J}, \ref{supp-table:mpcp-by-w}, and \ref{supp-table:computation-time-by-I}.
\begin{table}[ht]
\begin{center}
\begin{adjustbox}{width=0.8\textwidth}
\begin{tabular}{lrrrrrr}
\toprule
\multicolumn{1}{c}{\,} & \multicolumn{3}{c}{Gaussian} & \multicolumn{3}{c}{Binary}\\ \cmidrule(lr){2-4}\cmidrule(lr){5-7}
Basis Function & I=100 & I=500 & I=1000 & I=100 & I=500 & I=1000  \\ \midrule
Case 1  & 0.804 & 0.924 & 0.927 & 0.896 & 0.921 & 0.927\\
\midrule Case 2 & 0.836 & 0.929 & 0.941 & 0.911 & 0.935 & 0.939\\
\bottomrule
\end{tabular}
\end{adjustbox}
\end{center}
\caption{Mean pointwise coverage probability (MPCP) across $100$ repeated experiments. We fixed the number of visits $J = 10$, number of sampling points $L = 100$, and percentage of data used for constructing local bins $W = 5\%$.}
\label{table:mpcp-by-I}
\end{table}

Consider now simulations with missing data. Table \ref{table:mpcp-by-M} displays the MPCP statistics  during periods of missingness for Gaussian and binary data using both types of basis functions for different rates of missingness, $M$. Specifically, we calculated coverage probabilities at each domain point \emph{within each missing region} and then averaged the coverage probabilities across subjects, visits, and missing regions. The MPCP statistics are based on missing regions only, whose location and length vary among subjects and visits. Results are reported for $I = 1{,}000$, $J = 10$, $L = 100$, and  $W = 5\%$. Results in Table \ref{table:mpcp-by-M} indicate that the coverage probability of FRIM is robust to missing data even in regions where data are missing. As the missing rate $M$ increases, coverage probabilities increase slightly, but is close to $0.95$ and do not drift to $1$ even when missingness is substantial.

\begin{table}[ht]
\begin{center}
\begin{adjustbox}{width=0.8\textwidth}
\begin{tabular}{lrrrrrr}
\toprule
\multicolumn{1}{c}{\,} & \multicolumn{3}{c}{Gaussian} & \multicolumn{3}{c}{Binary}\\ \cmidrule(lr){2-4}\cmidrule(lr){5-7}
Basis Function & M=0.1 & M=0.25 & M=0.5 & M=0.1 & M=0.25 & M=0.5  \\ \midrule
Case 1  &  0.937 & 0.934 & 0.945 & 0.927 & 0.927 & 0.933\\
\midrule Case 2 & 0.927 & 0.932 & 0.948 & 0.939  & 0.939 & 0.937\\
\bottomrule
\end{tabular}
\end{adjustbox}
\end{center}
\caption{Mean pointwise coverage probability (MPCP) within missing regions. We fixed the number of subjects $I = 1,000$, number of visits $J = 10$, number of sampling points $L = 100$, and percentage of data used for constructing local bins $W = 5\%$.}
\label{table:mpcp-by-M}
\end{table}

\section{Application}\label{sec:app}
\subsection{Data Overview}
NHANES is a large-scale, ongoing study conducted in two-year waves by the National Center for Health Statistics, a unit of the Centers for Disease Control and Prevention in the United States. The study collects a wide range of health-related data from a nationally representative sample at each wave to assess the health and nutritional status of the United States population. Specifically, physical activity monitors (accelerometers) were deployed in NHANES 2003-2006 \citep{nhanes2003} and NHANES 2011-2014 \citep{nhanes2011, nhanes2013}, providing objective measurements of physical activity intensity with high resolution. 

For this application, we focus on the accelerometry data collected in the NHANES 2011-2014 study, which used wrist-worn accelerometers. All NHANES 2011-2014 participants aged 3 years and older were invited to wear a wrist-worn accelerometer (Actigraph GT3X+) for up to nine consecutive days, and $14{,}693$ individuals agreed to participate. Raw tri-axial acceleration data were recorded at $80$ Hz, and were extracted, processed, and released by the NHANES study team in Monitor Independent Movement Summary (MIMS) units \citep{john2019mims} at the minute level. A higher MIMS value indicates higher physical activity intensity. A logarithm transformation $f(x) = \log (1 + x)$ was applied to MIMS at each minute to reduce the data skewness, as suggested by \cite{varma2018total, cui2021additive}. 

To ensure data quality, we excluded days with less than $50$\% of estimated wear time, as labeled by the NHANES team. This threshold is far less stringent than the $95$\% threshold proposed by \citep{leroux2024nhanes}. By using this more lenient criterion, we preserved $11.2$\% more data that would have otherwise been excluded. This is possible because FRIM can handle missing data, as shown in the simulation experiments. We excluded individuals who had fewer than three days of wear and were younger than $60$ at the time of the study. The domain was down-sampled at 10-minute intervals so that each subject has $144$ observations per day. The resulting analytic sample consists of $2,285$ participants, $15,500$ days of wear, and $2,247,500$ observations ($2,202,076$ non-missing).

The primary question of interest is to quantify the uncertainty of the subject- and subject-visit-specific functional random effect trajectories, which is important both where data are observed and where it is missing. We applied FRIM using age, gender, and day of the week as fixed effects and set the bin width to $40$ minutes to balance estimation accuracy and computational cost. For each level, we kept $10$ eigenfunctions, which explained $92$\% and $62$\% percent of variations at each level, respectively. Inference was based on $2,000$ posterior samples after warm-up.

\subsection{Results}

\begin{figure}[!tbh]
  \centering
  \includegraphics[width=\textwidth]{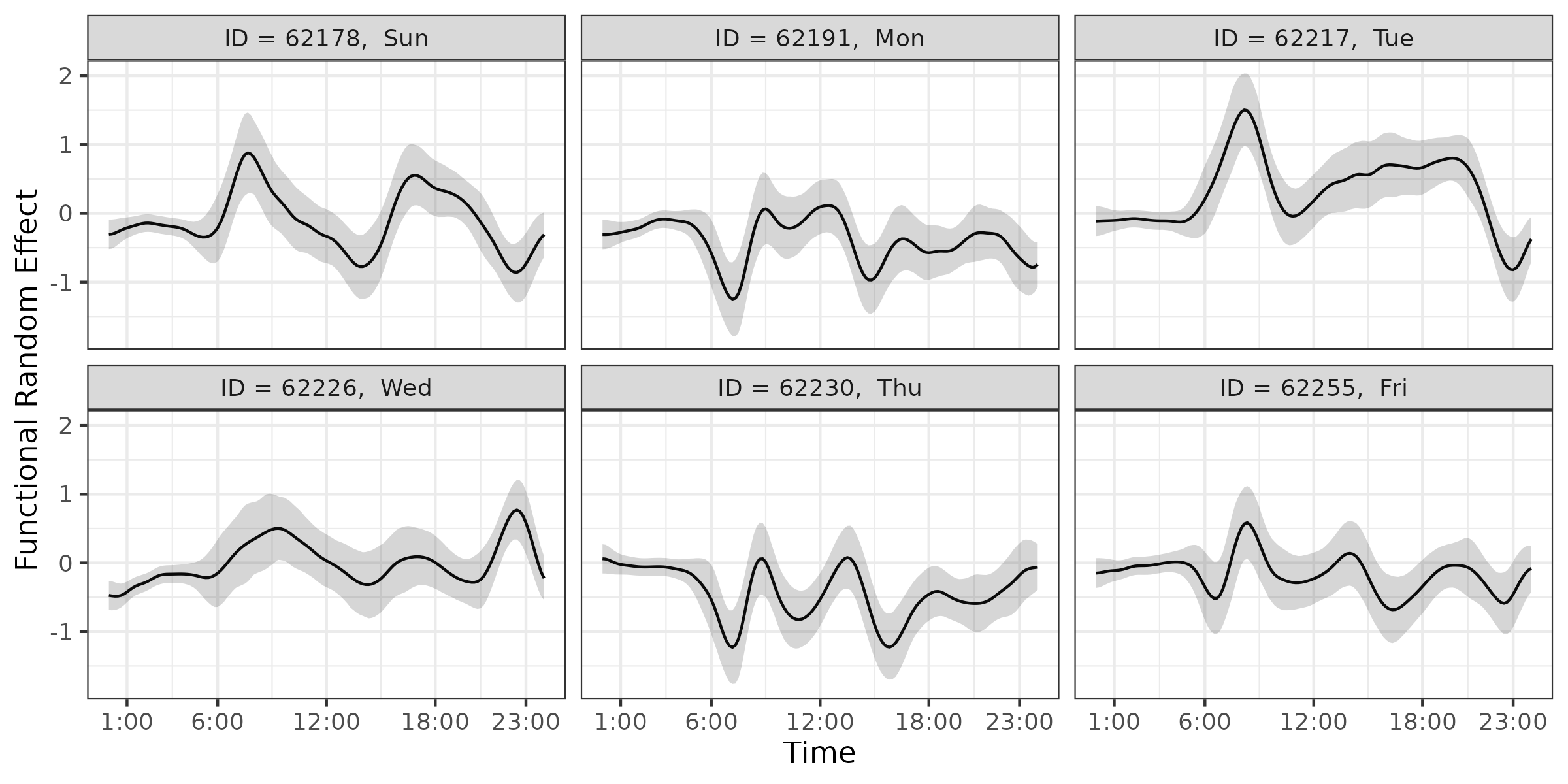}
  \caption{FRIM random effects predictions and 95\% credible intervals for six randomly selected visits from subjects without missing observations.}
  \label{fig:nhanes-nomissing}
\end{figure}

Figure \ref{fig:nhanes-nomissing} shows the subject-visit-specific random effect predictions using FRIM for six randomly selected subjects and visits with no missing observations. A distinctive feature of FRIM is its ability to quantify the uncertainty associated with these predictions, depicted by the gray ribbons representing pointwise $95\%$ credible intervals. The widths of the credible intervals vary depending on the variability of the observed data in the corresponding region. For example, all study participants tend to have narrower credible intervals at night when activity intensity variation is low and wider credible intervals during midday when activity is more variable. 

\begin{figure}[!tbh]
  \centering
  \includegraphics[width=0.8\textwidth]{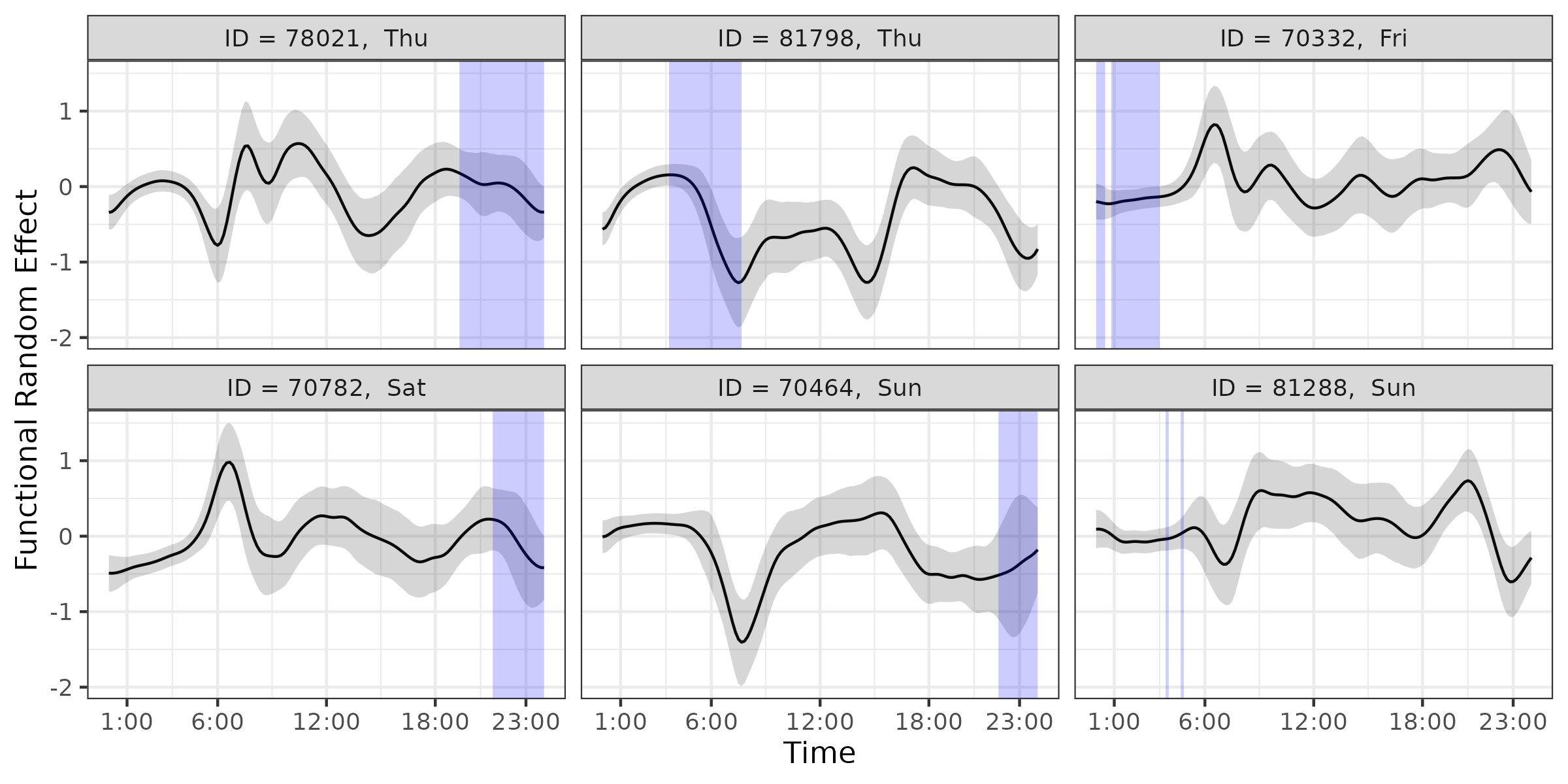}
  \caption{FRIM random effects predictions and 95\% credible bands for six randomly selected visits from subjects with missing observations. The missing regions are highlighted using blue boxes.}
  \label{fig:nhanes-missing}
\end{figure}

We also evaluated the performance of FRIM for study participants with missing data. Among the $1,775$ days containing missing data, we randomly selected six days displayed in Figure \ref{fig:nhanes-missing}. Missing regions are highlighted in blue and tend to appear as contiguous time intervals. This is somewhat expected, as missing data often result from battery depletion or device non-wear. The timing and duration of missingness vary across subjects and visits. Despite these variations, FRIM achieves reasonable predictions and $95\%$ credible intervals for all days, as the credible intervals do not increase substantially in areas with missing data. This is likely because FRIM can borrow information from the other days of the same individual and from other individuals.

We have also investigated the potential of FRIM for personalized anomaly detection. That is, given data for an individual over several days, could one quantify for this individual when and where new measurements are unusual? Specifically, given data from the first $J$ days of a study participant, we would like to check whether the estimated day $J+1$ level-two functional random effect falls within the $95$\% credible bands based on the first $J$ days. This problem has significant implications for personalized health monitoring, as a sudden deviation from an individual’s typical range could be associated with health outcomes. This provides a tailored and effective method for identifying unusual patterns for a specific individual. While we have not investigated the properties of FRIM for anomaly detection in simulation studies, the method is sensible and could be explored in depth in future research.

As a proof of concept, we focus on weekday activity from our analytic sample of $2{,}285$ study participants. The reason for excluding weekends is because of the extensive published evidence that weekday and weekend activity follow different patterns \citep{fda_with_R}. We further excluded $205$ participants with fewer than five weekdays of data to ensure that at least four days are available for training. The resulting analytic sample for this analysis consists of $2{,}080$ participants, $10{,}400$ days of wear, and $1{,}508{,}000$ observations, out of which $1{,}485{,}727$ are not missing. We randomly picked fours days from each subject for training and used the remaining day for anomaly detection. Specifically, training involves estimating eigenfunctions and constructing $95$\% credible bands for the day-specific functional random effects $b_{ij}(s)$ ($j \in$ training days). Training days are assumed to be exchangeable and we pooled the posterior samples from different days to construct the $95$\% credible bands. After training, we condition on the eigenfunctions estimated from the training data to sample scores for day $j'$ ($j'$ = test day) and, implicitly, the posterior samples for $b_{ij'}(s)$. The posterior mean of $b_{ij'}(s)$ is calculated and compared to the $95$\% credible bands constructed based on the training days. If the posterior mean of $b_{ij'}(s)$ falls outside the $95$\% credible bands for $s \in [s_l, s_u]$, this period is flagged for having potentially unusual activity level.

\begin{figure}[!tbh]
  \centering
  \includegraphics[width=0.8\textwidth]{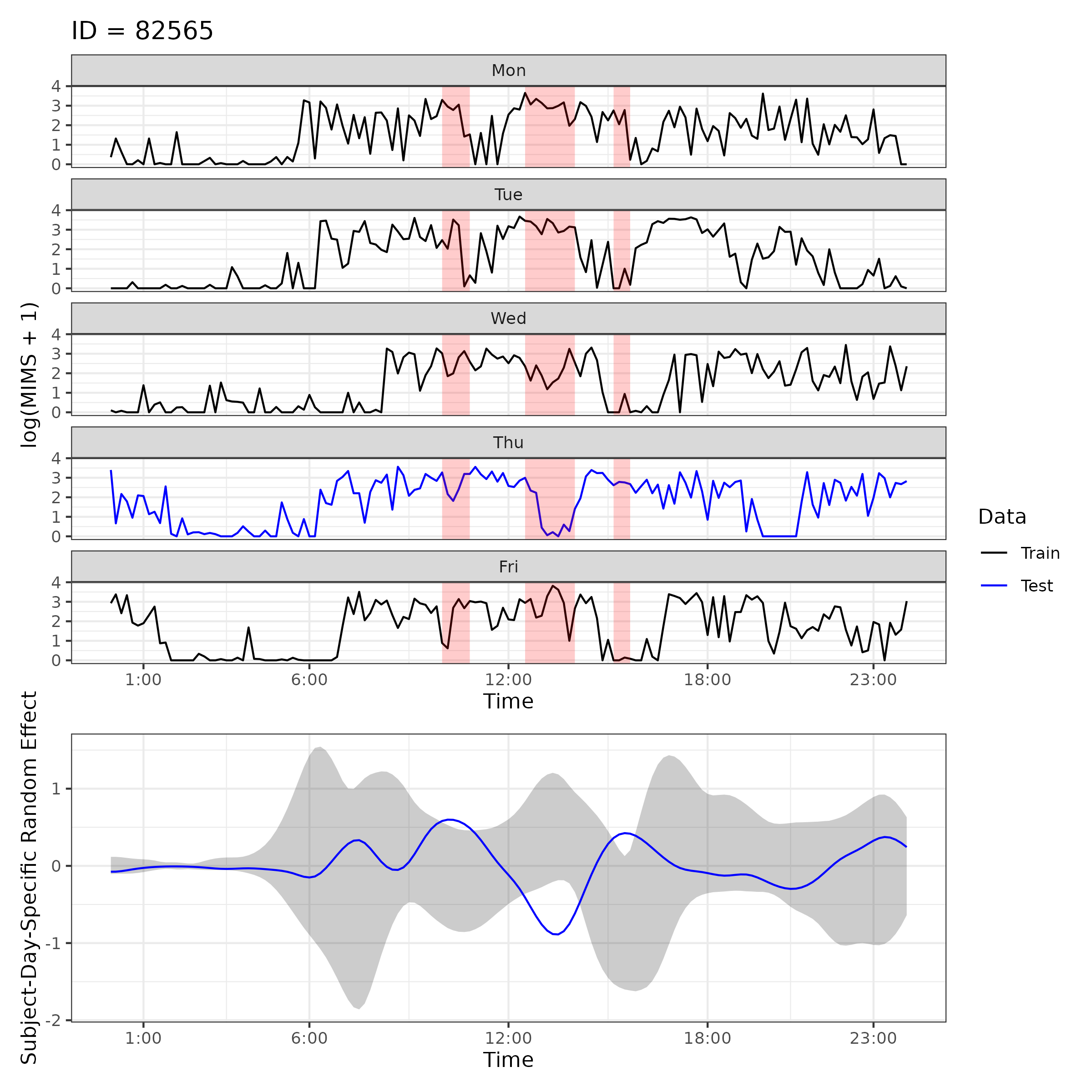}
  \caption{Personalized physical activity anomaly detection using FRIM for one NHANES accelerometry study participant. The top panel illustrates weekday physical activity for the study participant (black lines: training, blue lines: testing). Red boxes mark periods identified as anomalous. The bottom panel displays the day-specific (level-two) functional random effect. The gray ribbon represents the $95$\% credible interval derived from the training days, while the blue line indicates the point estimate for the test day.}
  \label{fig:nhanes-anomaly-detection}
\end{figure}

Out of $2,080$ participants, $38$ exhibited anomaly periods lasting longer than three hours. Figure \ref{fig:nhanes-anomaly-detection} presents results for one such individual. The top panel displays the log-transformed MIMS values on weekdays, with training days displayed in black and the test day in blue. Red boxes identify the time intervals where anomalies were detected. In the bottom panel, the gray ribbon represents the $95$\% credible intervals for the level-two (subject-visit-specific) functional random effect, constructed from the posterior samples based on the training days. The blue line displays the posterior mean of the level-two functional random effect for the test day.

The blue line deviates significantly from the $95$\% credible bands between 12:30 PM and 2 PM, corresponding to an unusually low activity level compared to the other days, as seen in the top panel. Additionally, minor deviations are observed around 10 AM and 3 PM, which correspond to subtler differences in activity levels observed in the top panel. More work is needed to evaluate the performance of this approach and the validity of model assumptions as well as the sample size needed to identify unusual days or time periods within a day, which is well beyond the scope of this work. Nonetheless, as a proof-of-concept, this approach demonstrates promise and warrants further exploration.

\section{Discussion}\label{sec:discussion}
We introduced an inference framework for correlated functional data to flexibly quantify the variability of subject- and subject-visit-specific trajectories. While functional mixed models have been widely studied in the literature, remarkably little is known about the performance of statistical inference for functional random effects. Our work bridges this gap through a combination of a local GLMM learning of the random effects structure and a global Bayesian procedure conditioning on the estimated residual structure modeled as structured functional random effects. Extensive simulations demonstrate that the proposed method achieves nominal coverage of the functional random effects while maintaining computational efficiency. FRIM is implemented in a reproducible software package (\url{https://github.com/xinkai-zhou/frim}) that enables efficient subject-specific predictions and inferences in large-scale studies.

\section{Acknowledgment}
The authors gratefully acknowledge use of the facilities at the Joint High Performance Computing Exchange (JHPCE) in the Department of Biostatistics, Johns Hopkins Bloomberg School of Public Health that have contributed to the research results reported within this paper.

\section{Supplementary Materials}

\subsection{Effect of number of visits}
Table \ref{supp-table:mpcp-by-J}  displays the MPCP statistics for Gaussian and binary data using both types of basis functions and at different number of visits $J$. In this experiment, we fixed the number of subjects $I = 1000$, number of sampling points $L = 100$, and the percentage of data used for constructing local bins $w = 5\%$. We observe that for Gaussian data, the number of visits plays a minor role in affecting coverage rate for either type of basis functions. For binary data, however, having just two visits is likely to impact the coverage rate. 
\begin{table}[ht]
\begin{center}
\begin{adjustbox}{width=0.8\textwidth}
\begin{tabular}{lrrrrrr}
\toprule
\multicolumn{1}{c}{\,} & \multicolumn{3}{c}{Gaussian} & \multicolumn{3}{c}{Binary}\\ \cmidrule(lr){2-4}\cmidrule(lr){5-7}
Basis Function & J=2 & J=5 & J=10 & J=2 & J=5 & J=10 \\ \midrule
Case 1  & 0.931 & 0.937 & 0.927 & 0.882 & 0.924 & 0.927\\
\midrule Case 2 & 0.942 & 0.932 & 0.941 & 0.888 & 0.921 & 0.939\\
\bottomrule
\end{tabular}
\end{adjustbox}
\end{center}
\caption{Mean pointwise coverage probability (MPCP) across $100$ repeated experiments at different number of visits. We fixed the number of subjects $I = 1000$, number of sampling points $L = 100$, and percentage of data used for constructing local bins $w = 5\%$.}
\label{supp-table:mpcp-by-J}
\end{table}

\subsection{Effect of bin width}
Table \ref{supp-table:mpcp-by-w}  displays the MPCP statistics for Gaussian and binary data using both types of basis functions and at different bin widths. In this experiment, we fixed the number of subjects $I = 1000$, number of visits $J = 10$, and number of sampling points $L = 100$. We observe that the overall coverage rate is satisfactory across all bin widths, with a bin width of 5\ providing a good coverage rate for both data types and basis functions.
\begin{table}[ht]
\begin{center}
\begin{adjustbox}{width=0.8\textwidth}
\begin{tabular}{lrrrrrr}
\toprule
\multicolumn{1}{c}{\,} & \multicolumn{3}{c}{Gaussian} & \multicolumn{3}{c}{Binary}\\ \cmidrule(lr){2-4}\cmidrule(lr){5-7}
Basis Function & w=2\% & w=5\% & w=10\% & w=2\% & w=5\% & w=10\%  \\ \midrule
Case 1  & 0.879 & 0.927 & 0.890 & 0.917 & 0.927 & 0.920\\
\midrule Case 2 & 0.919 & 0.941 & 0.934 & 0.876 & 0.939 & 0.941\\
\bottomrule
\end{tabular}
\end{adjustbox}
\end{center}
\caption{Mean pointwise coverage probability (MPCP) across $100$ repeated experiments at different bin widths. We fixed the number of subjects $I=1000$, number of visits $J = 10$, number of sampling points $L = 100$.}
\label{supp-table:mpcp-by-w}
\end{table}

\subsection{Computation time}
Table \ref{supp-table:computation-time-by-I} displays the median computation time in hours across 100 simulations at different sample sizes. The median computation time increases roughly linearly with the number of subjects $I$. 
\begin{table}[ht]
\begin{center}
\begin{adjustbox}{width=0.8\textwidth}
\begin{tabular}{lrrrrrr}
\toprule
\multicolumn{1}{c}{\,} & \multicolumn{3}{c}{Gaussian} & \multicolumn{3}{c}{Binary}\\ \cmidrule(lr){2-4}\cmidrule(lr){5-7}
Basis Function & I=100 & I=500 & I=1000 & I=100 & I=500 & I=1000  \\ \midrule
Case 1 & 2.0 & 6.6 & 14.6 & 1.1 & 5.1 & 11.4\\
\midrule Case 2 & 2.9 & 12.8 & 39.2 & 1.0 & 6.9 & 13.5\\
\bottomrule
\end{tabular}
\end{adjustbox}
\end{center}
\caption{Median computation time (hours) across $100$ repeated experiments at different sample sizes. In this experiment, we fixed the number of visits $J = 10$ and percentage of data used for constructing local bins $w = 5\%$. The number of MCMC warm-up and post warm-up iterations were set to $1,000$ and $2,000$, respectively.}
\label{supp-table:computation-time-by-I}
\end{table}
\bigskip

\clearpage\pagebreak\newpage
\baselineskip=14pt
\bibliographystyle{apalike}
\bibliography{References}

\begin{thebibliography}{}

\bibitem[Alam and Staicu, 2024]{alam_modeling_2024}
Alam, M.~S. and Staicu, A.-M. (2024).
\newblock Modeling longitudinal skewed functional data.
\newblock {\em Biometrics}, 80(4).

\bibitem[Brockhaus et~al., 2015]{brockhaus2015functional}
Brockhaus, S., Scheipl, F., Hothorn, T., and Greven, S. (2015).
\newblock The functional linear array model.
\newblock {\em Statistical Modelling}, 15(3):279--300.

\bibitem[Burbano-Moreno and Mayrink, 2024]{burbano-moreno_spatial_2024}
Burbano-Moreno, A.~A. and Mayrink, V.~D. (2024).
\newblock Spatial functional data analysis: irregular spacing and {Bernstein}
  polynomials.
\newblock {\em Spatial Statistics}, 60:100832.

\bibitem[Cao et~al., 2024]{cao2024efficient}
Cao, C., Cao, J., Pan, H., Zhang, Y., Jiang, F., and Li, X. (2024).
\newblock An efficient two-dimensional functional mixed-effect model framework
  for repeatedly measured functional data.
\newblock {\em arXiv preprint arXiv:2409.03296}.

\bibitem[Cao et~al., 2018]{cao2018robust}
Cao, C., Shi, J.~Q., and Lee, Y. (2018).
\newblock Robust functional regression model for marginal mean and
  subject-specific inferences.
\newblock {\em Statistical Methods in Medical Research}, 27(11):3236--3254.

\bibitem[Carpenter et~al., 2017]{carpenter2017stan}
Carpenter, B., Gelman, A., Hoffman, M.~D., Lee, D., Goodrich, B., Betancourt,
  M., Brubaker, M.~A., Guo, J., Li, P., and Riddell, A. (2017).
\newblock {STAN}: a probabilistic programming language.
\newblock {\em Journal of Statistical Software}, 76.

\bibitem[Crainiceanu et~al., 2024]{fda_with_R}
Crainiceanu, C.~M., Goldsmith, J., Leroux, A., and Cui, E. (2024).
\newblock {\em Functional Data Analysis with R}.
\newblock Chapman and Hall/CRC.

\bibitem[Cui et~al., 2021]{cui2021additive}
Cui, E., Crainiceanu, C.~M., and Leroux, A. (2021).
\newblock Additive functional cox model.
\newblock {\em Journal of Computational and Graphical Statistics},
  30(3):780--793.

\bibitem[Cui et~al., 2022]{cui2022fui}
Cui, E., Leroux, A., Smirnova, E., and Crainiceanu, C.~M. (2022).
\newblock Fast univariate inference for longitudinal functional models.
\newblock {\em Journal of Computational and Graphical Statistics},
  31(1):219--230.

\bibitem[Cui et~al., 2023]{cui2023mfpca}
Cui, E., Li, R., Crainiceanu, C.~M., and Xiao, L. (2023).
\newblock Fast multilevel functional principal component analysis.
\newblock {\em Journal of Computational and Graphical Statistics},
  32(2):366--377.

\bibitem[Di et~al., 2009]{di2009multilevel}
Di, C.-Z., Crainiceanu, C.~M., Caffo, B.~S., and Punjabi, N.~M. (2009).
\newblock Multilevel functional principal component analysis.
\newblock {\em The Annals of Applied Statistics}, 3(1):458.

\bibitem[Goldsmith et~al., 2011]{goldsmith_penalized_2011}
Goldsmith, J., Bobb, J., Crainiceanu, C.~M., Caffo, B., and Reich, D. (2011).
\newblock Penalized functional regression.
\newblock {\em Journal of Computational and Graphical Statistics},
  20(4):830--851.

\bibitem[Goldsmith et~al., 2015]{goldsmith2015generalized}
Goldsmith, J., Zipunnikov, V., and Schrack, J. (2015).
\newblock Generalized multilevel function-on-scalar regression and principal
  component analysis.
\newblock {\em Biometrics}, 71(2):344--353.

\bibitem[Greven et~al., 2011]{greven2011longitudinal}
Greven, S., Crainiceanu, C., Caffo, B., and Reich, D. (2011).
\newblock Longitudinal functional principal component analysis.
\newblock In {\em Recent Advances in Functional Data Analysis and Related
  Topics}, pages 149--154. Springer.

\bibitem[Greven and Scheipl, 2017]{greven_general_2017}
Greven, S. and Scheipl, F. (2017).
\newblock A general framework for functional regression modelling.
\newblock {\em Statistical Modelling}, 17(1-2):1--35.

\bibitem[Gunning et~al., 2024]{gunning2024mmlf}
Gunning, E., Golovkine, S., Simpkin, A.~J., Burke, A., Dillon, S., Gore, S.,
  Moran, K., O'Connor, S., Whyte, E., and Bargary, N. (2024).
\newblock A multivariate multilevel longitudinal functional model for
  repeatedly observed human movement data.

\bibitem[Guo, 2002]{guo2002functional}
Guo, W. (2002).
\newblock Functional mixed effects models.
\newblock {\em Biometrics}, 58(1):121--128.

\bibitem[John et~al., 2019]{john2019mims}
John, D., Tang, Q., Albinali, F., and Intille, S. (2019).
\newblock An open-source monitor-independent movement summary for accelerometer
  data processing.
\newblock {\em Journal for the Measurement of Physical Behaviour}, 2(4):268 --
  281.

\bibitem[Koner and Staicu, 2023]{koner_second-generation_2023}
Koner, S. and Staicu, A.-M. (2023).
\newblock Second-generation functional data.
\newblock {\em Annual Review of Statistics and Its Application}, 10(Volume 10,
  2023):547--572.

\bibitem[Leroux et~al., 2023]{leroux2023gfpca}
Leroux, A., Crainiceanu, C., and Wrobel, J. (2023).
\newblock Fast generalized functional principal components analysis.
\newblock {\em arXiv preprint arXiv:2305.02389}.

\bibitem[Leroux et~al., 2024]{leroux2024nhanes}
Leroux, A., Cui, E., Smirnova, E., Muschelli, J., Schrack, J.~A., and
  Crainiceanu, C.~M. (2024).
\newblock {NHANES} 2011-2014: Objective physical activity is the strongest
  predictor of all-cause mortality.
\newblock {\em Medicine and Science in Sports and Exercise}.

\bibitem[Li et~al., 2021]{li2021multilevel}
Li, Y., Nguyen, D.~V., Banerjee, S., Rhee, C.~M., Kalantar-Zadeh, K.,
  K{\"u}r{\"u}m, E., and {\c{S}}ent{\"u}rk, D. (2021).
\newblock Multilevel modeling of spatially nested functional data:
  Spatiotemporal patterns of hospitalization rates in the {US} dialysis
  population.
\newblock {\em Statistics in Medicine}, 40(17):3937--3952.

\bibitem[Lin et~al., 2024]{lin2024multilevel}
Lin, W., Zou, J., Di, C., Rock, C.~L., and Natarajan, L. (2024).
\newblock Multilevel longitudinal functional principal component model.
\newblock {\em Statistics in Medicine}, 43(25):4781--4795.

\bibitem[Lu et~al., 2024]{lu2024gcfpca}
Lu, Y., Zhou, X., Cui, E., Rogers, D., Crainiceanu, C.~M., Wrobel, J., and
  Leroux, A. (2024).
\newblock Generalized conditional functional principal component analysis.
\newblock {\em arXiv preprint arXiv:2411.10312}.

\bibitem[Marco et~al., 2024]{marco_functional_2024}
Marco, N., Şentürk, D., Jeste, S., DiStefano, C., Dickinson, A., and Telesca,
  D. (2024).
\newblock Functional mixed membership models.
\newblock {\em Journal of Computational and Graphical Statistics},
  33(4):1139--1149.

\bibitem[Morris and Carroll, 2006]{morris2006wavelet}
Morris, J.~S. and Carroll, R.~J. (2006).
\newblock Wavelet-based functional mixed models.
\newblock {\em Journal of the Royal Statistical Society Series B: Statistical
  Methodology}, 68(2):179--199.

\bibitem[NCHS, 2006]{nhanes2003}
NCHS (2006).
\newblock National health and nutrition examination survey data.
\newblock
  \url{https://wwwn.cdc.gov/nchs/nhanes/continuousnhanes/default.aspx?BeginYear=2003}.
\newblock (Accessed Dec 16, 2024).

\bibitem[NCHS, 2012]{nhanes2011}
NCHS (2012).
\newblock National health and nutrition examination survey data.
\newblock
  \url{https://wwwn.cdc.gov/nchs/nhanes/continuousnhanes/default.aspx?BeginYear=2011}.
\newblock (Accessed Dec 16, 2024).

\bibitem[NCHS, 2014]{nhanes2013}
NCHS (2014).
\newblock National health and nutrition examination survey data.
\newblock
  \url{https://wwwn.cdc.gov/nchs/nhanes/continuousnhanes/default.aspx?BeginYear=2013}.
\newblock (Accessed Dec 16, 2024).

\bibitem[Park and Staicu, 2015]{park_longitudinal_2015}
Park, S.~Y. and Staicu, A.-M. (2015).
\newblock Longitudinal functional data analysis.
\newblock {\em Stat (International Statistical Institute)}, 4(1):212--226.

\bibitem[Qian et~al., 2024]{qian2024multivariate}
Qian, Q., Nguyen, D.~V., Telesca, D., Kurum, E., Rhee, C.~M., Banerjee, S., Li,
  Y., and Senturk, D. (2024).
\newblock Multivariate spatiotemporal functional principal component analysis
  for modeling hospitalization and mortality rates in the dialysis population.
\newblock {\em Biostatistics}, 25(3):718--735.

\bibitem[Scheipl et~al., 2016]{scheipl2016generalized}
Scheipl, F., Gertheiss, J., and Greven, S. (2016).
\newblock Generalized functional additive mixed models.
\newblock {\em Electronic Journal of Statistics}, pages 1455--1492.

\bibitem[Scheipl et~al., 2015]{scheipl2015functional}
Scheipl, F., Staicu, A.-M., and Greven, S. (2015).
\newblock Functional additive mixed models.
\newblock {\em Journal of Computational and Graphical Statistics},
  24(2):477--501.

\bibitem[Serban et~al., 2013]{serban_multilevel_2013}
Serban, N., Staicu, A.-M., and Carroll, R.~J. (2013).
\newblock Multilevel cross-dependent binary longitudinal data.
\newblock {\em Biometrics}, 69(4):903--913.

\bibitem[Sergazinov et~al., 2023]{sergazinov2023case}
Sergazinov, R., Leroux, A., Cui, E., Crainiceanu, C., Aurora, R.~N., Punjabi,
  N.~M., and Gaynanova, I. (2023).
\newblock A case study of glucose levels during sleep using multilevel fast
  function on scalar regression inference.
\newblock {\em Biometrics}.

\bibitem[Shamshoian et~al., 2020]{shamshoian_bayesian_2020}
Shamshoian, J., Şentürk, D., Jeste, S., and Telesca, D. (2020).
\newblock Bayesian analysis of longitudinal and multidimensional functional
  data.
\newblock {\em Biostatistics}, 23(2):558--573.

\bibitem[Shou et~al., 2015]{shou2015structured}
Shou, H., Zipunnikov, V., Crainiceanu, C.~M., and Greven, S. (2015).
\newblock Structured functional principal component analysis.
\newblock {\em Biometrics}, 71(1):247--257.

\bibitem[Staicu et~al., 2010]{staicu2010fast}
Staicu, A.-M., Crainiceanu, C.~M., and Carroll, R.~J. (2010).
\newblock Fast methods for spatially correlated multilevel functional data.
\newblock {\em Biostatistics}, 11(2):177--194.

\bibitem[Sun and Kowal, 2024]{sun2023ultra}
Sun, T.~Y. and Kowal, D.~R. (2024).
\newblock Ultra-efficient {MCMC} for {B}ayesian longitudinal functional data
  analysis.
\newblock {\em Journal of Computational and Graphical Statistics}, 0(0):1--13.

\bibitem[Varma et~al., 2018]{varma2018total}
Varma, V.~R., Dey, D., Leroux, A., Di, J., Urbanek, J., Xiao, L., and
  Zipunnikov, V. (2018).
\newblock Total volume of physical activity: {TAC}, {TLAC} or {TAC}
  ($\lambda$).
\newblock {\em Preventive Medicine}, 106:233--235.

\bibitem[Volkmann et~al., 2023]{volkmann_multivariate_2023}
Volkmann, A., Stöcker, A., Scheipl, F., and Greven, S. (2023).
\newblock Multivariate functional additive mixed models.
\newblock {\em Statistical Modelling}, 23(4):303--326.

\bibitem[Xiao et~al., 2016]{xiao2016fast}
Xiao, L., Zipunnikov, V., Ruppert, D., and Crainiceanu, C. (2016).
\newblock Fast covariance estimation for high-dimensional functional data.
\newblock {\em Statistics and Computing}, 26:409--421.

\bibitem[Zhou et~al., 2023]{zhou2023gmfpca}
Zhou, X., Wrobel, J., Crainiceanu, C.~M., and Leroux, A. (2023).
\newblock Analysis of active/inactive patterns in the {NHANES} data using
  generalized multilevel functional principal component analysis.
\newblock {\em arXiv preprint arXiv:2311.14054}.

\bibitem[Zhu et~al., 2019]{zhu2019fmem}
Zhu, H., Chen, K., Luo, X., Yuan, Y., and Wang, J.-L. (2019).
\newblock {FMEM}: Functional mixed effects models for longitudinal functional
  responses.
\newblock {\em Statistica Sinica}, 29(4):2007.

\bibitem[Zipunnikov et~al., 2014]{zipunnikov2014longitudinal}
Zipunnikov, V., Greven, S., Shou, H., Caffo, B., Reich, D.~S., and Crainiceanu,
  C. (2014).
\newblock Longitudinal high-dimensional principal components analysis with
  application to diffusion tensor imaging of multiple sclerosis.
\newblock {\em The Annals of Applied Statistics}, 8(4):2175.

\end{thebibliography}

\end{document}